# QUALITY INDICATORS FOR COLLECTIVE SYSTEMS RESILIENCE

**Vincenzo De Florio, PATS research group,**
**Universiteit Antwerpen & iMinds research institute,**
**Middelheimlaan 1, 2020 Antwerpen-Berchem, Belgium.**
**Tel.: +32-3-2653905, fax: +32-3-2653777, e-mail: vincenzo.deflorio@uantwerpen.be**

## ABSTRACT

Resilience is widely recognized as an important design goal though it is one that seems to escape a general and consensual understanding. Often mixed up with other system attributes; traditionally used with different meanings in as many different disciplines; sought or applied through diverse approaches in various application domains, resilience in fact is a multi-attribute property that implies a number of constitutive abilities. To further complicate the matter, resilience is not an absolute property but rather it is the result of the match between a system, its current condition, and the environment it is set to operate in. In this paper we discuss this problem and provide a definition of resilience as a property measurable as a system-environment fit. This brings to the foreground the dynamic nature of resilience as well as its hard dependence on the context. A major problem becomes then that, being a dynamic figure, resilience cannot be assessed in absolute terms. As a way to partially overcome this obstacle, in this paper we provide a number of indicators of the quality of resilience. Our focus here is that of collective systems, namely those systems resulting from the union of multiple individual parts, sub-systems, or organs. Through several examples of such systems we observe how our indicators provide insight, at least in the cases at hand, on design flaws potentially affecting the efficiency of the resilience strategies. A number of conjectures are finally put forward to associate our indicators with factors affecting the quality of resilience.

## INTRODUCTION

Resilience is one of those concepts that seem to escape a general and consensual understanding. In fact resilience is often mixed up with other system attributes, e.g., robustness, elasticity, evolvability, performability, adaptivity, dependability, antifragility—to name but a few. Resilience is also one of those "overloaded" terms that throughout the years have been adopted, albeit assuming different meanings, in several domains and disciplines including, for instance, psychology, microeconomics, computer networks, security, safety, management science, biology, ecology, cybernetics, and control theory [TriDG09, SaMa11, Mey09, Lap05]. Applications resilience is often associated to are also many and diverse. They include for instance crisis management; ecosystems monitoring, preservation, and conservation; civil defense; business ecosystems; and many others [Car+12].

This paper focuses on the above problem and specifically on how to discuss about a system's quality of resilience. In order to do so we first briefly introduce, in Section "Resilience", the main subject of our discussion. In that section resilience is identified as a multi-attribute property combining evolvability with identity robustness. A major conclusion is also reached, namely that resilience should be recognized as a dynamic figure that cannot be assessed in absolute terms but only with respect to the currently experienced environmental conditions and the way those conditions evolve over time. Finally, resilience is characterized has having either an individual or a social dimension. Following section, "Limiting factors for individual strategies", elaborates on the latter aspect and provides a number of arguments to justify the fact that, at least in certain cases, the purely individualistic approach to resilience is not sufficient, while other social-oriented approaches (for instance those based on mutualistic relationships and other biologically-inspired collaborative interactions) may be more effective depending on the situation at hand.

In Section "Resilience Quality Indicators" we then propose an indirect approach to reasoning about quality of resilience. We do so by proposing three quality indicators: behavior, identity, and organization. We show how each of those indicators represents a different concern of resilience and provides a distinct viewpoint to the characteristics of an individual or a collective system's resilience. Here we also put forward our major conjecture, namely that a mismatch between the three indicators may signal a lock-in of a resilience strategy, i.e., a design factor preventing the strategy to fully achieve its intended goals.

After this, in Section "Discussion", we consider three examples of collective systems for the ambient assistance to the elderly and the impaired. In each case we describe the system and some of its major operational assumptions and we assess qualitatively the identity, behavior, and organizational aspects of those systems. Mismatches among the quality indicators and corresponding lock-ins are highlighted. A number of conjectures are exposed regarding the relation between those mismatches, the context, and the quality of resilience.

Major lessons learned and our next steps in this framework of problems are finally reported in Section "Conclusions".

# RESILIENCE

Resilience may be defined as a system's ability to absorb or tolerate change without losing their peculiar traits or expected behaviors [DF13a]. As can be readily understood, this definition consists of two separate parts corresponding to two distinct features. The first such feature is **evolvability**, namely the ability to "alter [one's] structure or function so as to adapt to changing circumstances" [Jen04]. The second feature is **identity robustness**, which is the ability of an evolving system to retain their features and characteristics in spite of the experienced perturbations or even of the adjustments operated in order to counterbalance changes. Thus if we consider, e.g., a hard real-time system, say $s$, and we subject it to some perturbation, we shall say that $s$ is resilient (*with respect to those perturbations*) if both the following two conditions hold:

1) $s$ does not experience crash failures.
2) $s$ does not experience performance failures (i.e., timing failures) affecting its nature of hard real-time system.

The above two conditions closely correspond to Aristotelian **entelechy**, discussed by the Great Scholar's in his Physics and Psychology [Sa1995, Laws1986] . Sachs' translation of entelechy is particularly intriguing in that it closely corresponds to the above two conditions: for Sachs entelechy is "being-at-work while staying-the-same" [Sa1995].As discussed in [DF13a], key prerequisites to resilience are the ability to perceive, become aware, and plan: as an example the ability to *react from change* requires the ability to *perceive change*, while the ability to *anticipate change* calls for the ability to *become aware* of trends and patterns. In the cited reference we show how the variety and quality of the services associated with those abilities have a strong impact on resilience. In particular the choice of the perception subsystems implies a hard coupling with a reference environment. Because of this coupling resilience always refers (implicitly or otherwise) to said reference environment: any real-life (namely, non-perfect) system is resilient *with respect to certain conditions*, and *not others*. As an example, it would make little sense to state that a man is "more resilient" than a dog in absolute terms. In fact, despite his superior features, a man cannot perceive sounds and lights outside certain ranges defined by his "design" (the characteristics of the human species) and his state (the characteristics of the given individual man). As an example, a man may not perceive certain noise waves that on the contrary are well within a dog's perception range. Thus a dog may become aware and possibly escape from a threat forewarned by, e.g., ultrasonic noise, while a man would be caught unawares by it.

The above definition also highlights how resilience may be reached through two options: either absorbing or tolerating change. In the first case this is reached by masking the consequences of change by means of physical or design redundancy [John1989, Avi1995]. We refer to this property as to elasticity. In the latter case, specific active behaviors must be enacted in order to counterbalance change. Said active behavior can be further differentiated into two main classes: individual-context and social-context resilient strategies [Eug09]. Individual-context strategies aim at reaching resilience from an individual perspective and with little or no interaction with neighboring entities. If present, said interaction is non-collaborative. On the contrary social-context strategies involve richer forms of interaction such as the ability to enact collective strategies or establish mutualistic relationships.

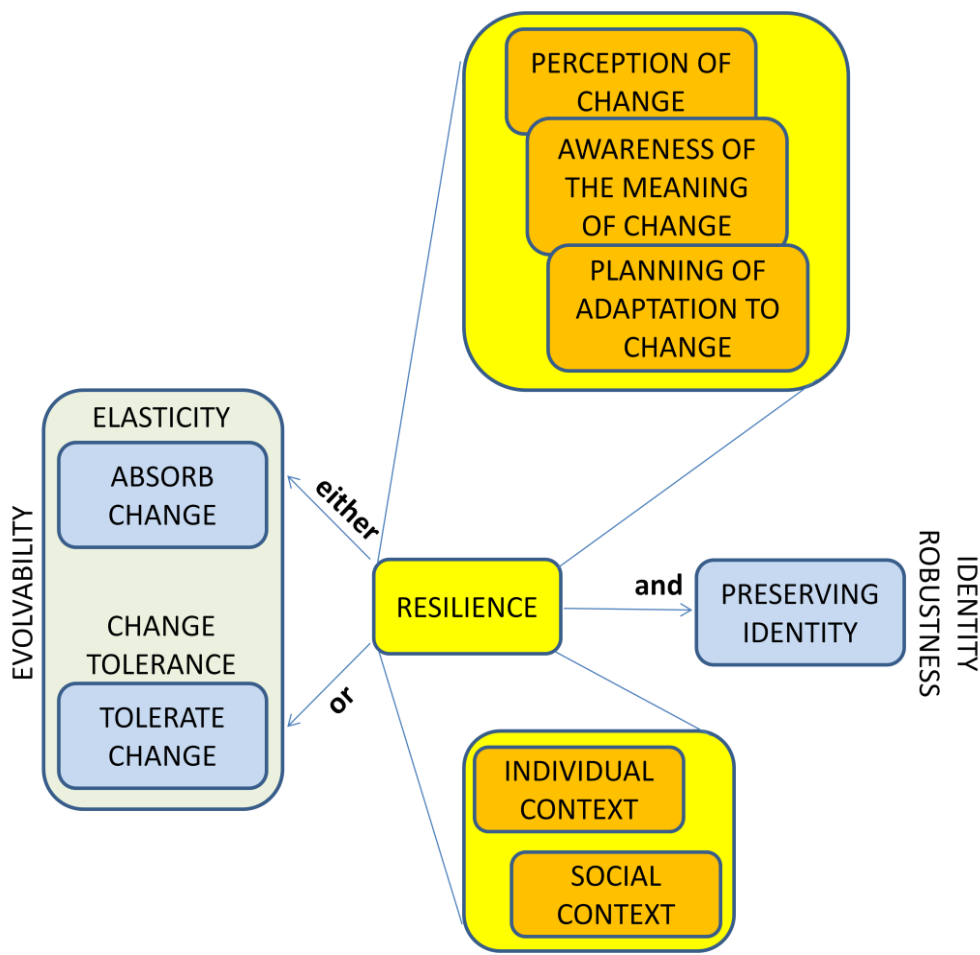

Figure 1. Various aspects of resilience are depicted: Resilience is identity preservation coupled with evolvability; the latter may be reached through elasticity or change tolerance; requirements for resilience include perception, awareness, and planning abilities; which may be enacted through individual-context and social context strategies.

We now focus on the just mentioned classes and highlight how resorting only on individual-oriented strategies may limit the quality of resilience strategies.

# LIMITING FACTORS IN INDIVIDUAL STRATEGIES

In previous section we highlighted a number of facts about resilience and in particular the availability of 1) individualistic strategies centered on a "focal entity" competing for resilience, and 2) social strategies in which various forms of mutualistic relationships are established. We put forward the conjecture that indeed both classes of strategies are necessary and in particular that the individual-context resilience strategies suffer from a number of limitations that can only be solved when extending those strategies so as to include the social context. A number of facts supporting our conjecture are summarized in what follows.

# LIMITED REACTIVE SPEED

We observe how reactive behaviors often operates on a time scale that is incompatible with change – especially in the face of turbulent and complex environments. Drastic and sudden changes call for prompt reaction, and a result of insufficient speed in the adaptive response is often the demise of the individual. Collaborative strategies may improve the quality of resilience, e.g., by artificially augmenting the perception range. In [DF13c] we showed this through an ambient intelligence scenario in which resilience could only be achieved through a collective perception strategy. Said strategy was inspired by the traditional use of canaries as sentinel species [vdS1999] to detect the presence of toxic gases in coal mines.

# CONSTRAINTS OF DESIGN

A second limitation derives from the typical constraints affecting system design. Due to technological constraints, physical limits, economic considerations, marketing strategies, and other reasons, the quality and the number of perception, awareness, planning, and executive organs is intrinsically limited. This leads to systems tailored towards specific reference conditions and environments. This system-environment association may be so pronounced that a system may be used as an "identifier" for the environment it is associated with – the term used in literature is "indicator species" [Farr02]. Again, mutualistic relationships based on collective perception may overcome design constraints.

Interestingly enough, a similar situation regards biological systems. Here the major limiting factors are physical limits and system organization. As already mentioned, resilience typically calls for different qualities—in the perception of the environment; the analysis of ongoing and past situations; the planning of strategies and reactions; as well as in their execution [DF13c]. Such qualities are a result of different "compartments" in a physical body, including e.g. the sensory apparatus, the endocrine, the nervous, the skeletal, and the muscular systems, all of which are organized into a single physical container—the body—and subjected to the same physical properties and limitations. An interesting example of this is reported in [Ni08, Ni10]: At some point in the history of biological evolution several species realized how "gathering more information from the environment improves chances of survival" [Ni08]. As a result, the sensory apparatus was privileged with respect to other design aspects, which translated in new species being endowed with a high number and variety of sensory receptors. On the other hand, gathering too large an amount of information calls for allocating too much of a system's room to its sensory pathways. This goes to the detriment of other abilities—including the ability to timely analyze and correlate the gathered information, comparing it with past patterns, and ultimately making use of it to better match the challenges posed by the environments. In other words the larger the amount of available sensors, the less physical room exists to accommodate *other* features and behaviors, including those very qualities required by resilience. Through the mechanisms of evolution this major bottleneck induced new *system organizations* intended to widen the design spectrum and allow more and more features to be "squeezed in"[2]. This resulted in improved "designs" that passed through the sieve of natural selection but nevertheless were still a result of trade-offs between physical space, quality, and accuracy [Ni10]. Intra-species and inter-species mutualistic relationships constitute nature's answer to this state of things.

---

[2] An interesting example of this innovation is reported in [Ni08, Ni10]: Rabbits are able to avoid dangerous obstacles (e.g. thorns) with remarkable precision and timing while running away from predators. This achievement is possible due to several reasons: The sensory apparatus of rabbits accounts for an average of 2,000 hair receptors per square centimeter of rabbit ear. This adds up to over 100,000 hair receptors. Connecting such a huge amount of sensors to the brain would require a number of dendrites (nerve fibers) simply too large to be compatible with the "architecture" of a rabbit's brain. Admitting 100,000 perception endpoints in the brain would inhibit processing. Nature deals with this bottleneck by linking each hair to 4 neurons instead of a single one. Somehow a combination of convergent and divergent neural pathways allows to produce perception with satisfactory accuracy and timeliness but without sacrificing "too much" system room—which would impact on the higher functions of the brain and ultimately on evolvability and survivability. A similar mechanism was observed in the human visual system [WTW1955].

SPECIALIZATION

A third argument against a purely individual-oriented approach to resilience comes from considering that the main focus of each individual as well as each species is a privileged "task environment" [AF1983], namely a subset of the "global environment" defined by the individual's natural niche, its sensory apparatus, and the events it experienced. This specialization allows the individual to focus its attention and action on a subset of the environmental "context", namely on the set of figures that the individual (or the species) estimated to be the most important factors for the definition of its strategies. Though effective when the estimation is correct, this approach is in general quite risky as it makes the individual blind to the onset of new interrelationships that have the potential to affect considerably the environmental conditions. The definition of novel "super-entities" such as the flock, the pack, or the ecosystem, allows the negative effects of individual specializations to be reduced if the perception of a threat by any individual is transmitted to the rest of the group. An example of this may be found in [DF13a].

DIVERSITY AND DISPARITY

Yet another argument is provided by the collective response that nature adopts when facing so-called "Black Swan" events [Tal12]. As already mentioned, no absolute assessment is possible for any resilience strategy. Nevertheless, practical considerations in nature and technology mandate that choices are made in order to select which natural traits or design templates will need to reappear in the next generation. This means that both nature and computer designers 1) match alternative solutions to the "typical" environmental conditions they observe; and (2) promote those solutions that appear to best match those conditions. An example of this phenomenon in natural system was the adoption of mineralized skeletons. Under stable environmental condition this solution is quite practical and effective as it confers protection against predators [Knoll]. As a result, it became widespread across different species. This general diffusion introduced a *weakness*: "skeleton formation requires more than the ability to precipitate minerals; precipitation must be carried out in a *controlled fashion* in *specific biological environments*" [Knoll]. This hidden dependence triggered the greatest extinction event ever—the so-called Great Dying, or Permian–Triassic extinction event—which about 252 million years ago led to the wiping of 96% of all marine species and 70% of terrestrial vertebrate species. A sudden and unexpected event—a Black Swan—turned instantly the tables of evolution. What was resilient became fragile and was swept away. What is worse, the more and the longer a solution is assessed as preferable, the more it affects disparity—namely inter-species morphological diversity [Mar06]. Thus, the rarest the black swan, the greater is the percentage of affected species.

Nature's answer to this conundrum is once again in the form of collective strategies. Through mechanisms such as mutation and the above mentioned task environments nature guarantees that, at any given time, not all species adopt the same "best-matching" solutions. At least a minority of species is enough diverse to survive in the common conditions and thrive in the rarest of conditions. When the Black Swan introduces those rarest conditions, the minority species become the survivors on top of which the ecosystem may begin its recovery.

Interestingly enough, the same "disparity syndrome" may affect computer-based systems:

- A solution becomes widespread (for instance a memory technology, a software library, a programming language, an operating system, or a search engine).
- The solution introduces a weakness: for instance, a dependence on a hidden assumption, or a "bug" depending on certain subtle and very rare environmental conditions.
- This translates in a common trigger, a single-point-of-multiple-failures: rare events may "turn on" the weakness and hit hard on all the systems that made use of the solution.

Computer viruses typically exploit lack of technological disparity to reach epidemic diffusion [EaKl10]. The Millennium Bug is a case where such epidemic diffusion though dreaded did not actually occur[3]. Also in this case a solution able to guarantee service resilience is sought through a collective and diverse response. In order to decrease the chance of correlated failures, multiple diverse replicas are executed in parallel [Avi85] or one after the other [RaXu1995].

---

[3] As society got closer and closer to the year 2000, the possible presence of a design fault in our software, the so-called Millennium Bug , became a nightmare that seemed to threaten all those crucial functions of our society today appointed to programs manipulating

## TECHNOLOGY

A final argument against the purely individual model is given by the rapid diffusion of new technologies such as the semantic web [FHHNS07], the internet-of-things [Fl10], pervasive computing, and cyber-physical societies [Zu10], which only exacerbate the speed of evolution of the environments. We deem that the introduction of such a large number of new players with unprecedented abilities to monitor, analyze, correlate the context, as well as to plan and execute autonomic and collective adaptation strategies, is making it more and more inconceivable to capture singlehandedly the complexity of these interactions and the consequences on their environments as implied by individual-context strategies.

We conclude this section by observing how nature's answer to the above limitations and to harsh environmental conditions is given by the definition of social organizations and the adoption of collective strategies, which has the following advantages: First, it allows individual views and experiences on task environments to be merged into a larger view; and secondly, it allows the intrinsic limitations of individual perception, awareness, and planning capabilities to be overcome through the combination of individual features. We also highlight once more the important role played in collective strategies by diversity/disparity and multiplicity: The former provide coverage against rare events with strong impact on system-environment fits; the latter is important in that it implies *parallel* perception, analysis, planning, and reaction capabilities. The sheer number of individuals can be in fact a first line of defense against external shocks, e.g. predators, or it can produce a more effective "composite prey"[4].

# RESILIENCE QUALITY INDICATORS

We conjecture that, in order to better understand and compare the effectiveness of collective strategies, three aspects need to be explicitly taken into account.

A first aspect is given by the quality of behaviors that the system constituents are able to exercise. These behaviors may range from an individual cell's simple reactions to external stimuli to the complex strategies of a business ecosystem involving a mixture of cooperation and competition (as e.g. in co-opetition [BN1998]). In Subsection "Behaviors" we classify the possible behaviors leveraging on classic work on cybernetics [RWB1943] and general systems theory [Bou1956] and on recent work on cyber-physical systems [Lee08], societies [Zu10], and communities [SDe10, SDe13].

A second aspect is related to the characteristics of the individual and of the social organization it is set to operate in. Even aggregations of simple individuals may give raise to robust social organizations exhibiting a certain degree of complex behaviors, as it is the case e.g. for the beehive [Mae1909] or even bacteria [ML12, SWBO09]. Furthermore, the interplay between the individual and the social "persona" may create centripetal and centrifugal behaviors ranging from full identification in the "greater self" to selfishness and opportunism. This aspect is discussed in Subsection "Individual and Collective Personae".

A third aspect is discussed in Subsection "Organizational and Control Structures of Social organization". There we first briefly recall the characteristics of the centralized organizations, hierarchies, heterarchies, holarchies, and fractal organizationsWe highlight in particular the differences in control and feedback

---

calendar dates, such us utilities, transportation, health care, communication, public administration, and so forth. Though very diverse, all those functions were potentially using the same design, thus resulting in a loss of disparity. Luckily the expected many and possibly crucial system failures due to this one application-level fault were not so many and not that crucial.

[4] This concept is nicely rendered in [DGV1972], which I translate here as follows: "One hundred hands make up our strength; one hundred eyes stand guard over us—you are but one. Now if you want you may go, or stay and join with us."

flows, as well as the different ways to concert action. Our conclusion is that each organizational structure determines a different blend of strengths and weaknesses—of capabilities and deficiencies—and therefore it constitutes another important aspect to understand an organization' ability to adapt to the highly complex dynamics of turbulent environments.

BEHAVIORS

We believe that a first concept necessary to discuss resilience in systems and organizations is given by behavior. In their classic work [RWB1943] the authors introduce the concept of "behavioristic study of natural events" and propose a classification of entities according to their behavior. In the cited reference behavior is defined as "any change of an entity with respect to its surroundings". The authors observe how entities may be classified according to peculiar characteristics of their behaviors and identify several behavioral classes. Here we extend this classification by including individual and collective behaviors typical of complex socio-technical organizations. The extended classification includes the following concepts:

- Passive behavior, which takes place when an entity changes its state only by receiving energy from an external source. A kicked ball does not produce the energy that sets it in motion—it simply receives that energy; therefore, it is characterized by passive behavior.
- Active behavior occurs when an entity "is the source of the output energy involved in a given particular reaction" [RWB1943].
- Purposeful active behavior is active change meant to attain a goal—for instance survival or economical profit. In this class of behaviors the output energy is exerted so as to move from a certain state into another one. Its opposite is purposeless (that is, random) active behavior. Purposeful active behavior is not restricted to beings but it may also pertain, e.g., to servo-mechanisms, cyber-physical systems, and legal persons.
- Teleological behavior is an entity's purposeful active behavior that is "controlled by the margin of error at which the [entity] stands at a given time with reference to a relatively specific goal". We observe how this form of behavior requires two key capabilities: *i*) being able to perceive the relationship between one's actions and one's goal, and *ii*) being able to adjust dynamically one's behavior so as to maximize the chances to reach one's goal. In [DF13c] we refer to the above two properties as to perception and apperception.
- Simple individual extrapolative (that is, predictive) behavior is one in which the acting entity possesses a third feature, namely *iii*) the ability to formulate its action in function of an extrapolated future state along a single or a few dimensions. This choice is done in isolation, i.e. without considering the choices of the entities co-existing in the same environment. An exemplary behavior in this class is speculation in programming language compilers: The compiler predicts the outcome of branch statements (in "IF's" and loops) to artificially enlarge the block of instructions to be pipelined for execution [HP06]. This prediction is done considering an individual piece of code at a time.
- In social predictive teleological behavior a fourth capability is added, namely *iv*) the ability to operate "quorum sensing": The extrapolated future state takes into account the current and the hypothesized future states of neighboring entities. Traders and trading software perceive the state of the stock exchange market; (try to) predict the predominant choice of the players and therefore the future state of the marked; and change their behaviors by using their predictions. Interestingly enough, these behaviors are *not* a prerogative of highly evolved species. Recent research revealed that such complex forms of behavior are also present in communities of very simple beings, such as the Bacillus subtilis bacterium. When subjected to a stressful environment (e.g. due to drought, radiation, or overpopulation [ML12, SWBO09]) these bacteria adopt quorum sensing and choose between cooperative and selfish strategies depending on an estimation of the choices made by their fellows—a sort of gigantic generalization of the Prisoner's Dilemma [Dr1981]. Another example is given by the strategies to prevent cache misses in multi-user operating systems. In such a context multiple users compete for the acquisition of a scarce resource—cache memory—and the operating system needs to base its strategies upon the cumulative action of all users [HP06].
- In complex multivariate extrapolative behavior, computing the future state also requires *v*) the ability to perform multiple extrapolations along different dimensions, e.g. a temporal and a spatial axis, on an individual or a social scale. An example of this behavior is given by the classical role of managers before the introduction of collective strategies and business ecosystems: Planning an organizational strategy speculating a future state in function of the multiple indications individually perceived and deducted from the environment. It is important to remark how the focus here is purely an endogenous one: It is "matching" [singlehandedly] "organizational capacities to environmental demands" [AF1983].

- A final class of behaviors is given by behaviors backed up by future-responsive social learning & collective strategies [Mi1973, DSMF75, AF1983] (called purposive anticipatory behaviors in [ECW05]). Such behaviors constitute the subject of human ecology: "*Collective and proactive* forms of organizational adaptation to the environment" requiring *vi*) the ability to foster the onset of "collectively constructed and controlled social environments" [AF1983] on top of the physical environment, in a way logically similar to the casting of overlay networks on top of the Internet. These dynamic *social overlay networks*, taking the shape of e.g. communities of mutual support [DFB10, SDe07, SDe10], may perform collective behaviors beyond cooperation and competition, e.g. co-opetition. In addition to the ability to extrapolate through many dimensions this class of behaviors is characterized by the ability to make choices with different planning horizons, e.g. initiating temporary alliances with natural competitors in order to reach the collective "social energy" [DFB10] necessary to exploit otherwise unreachable opportunities [GPT06]. This requires being able to estimate and reach the "critical social mass" required to modify the state of things and "create opportunities for future success the way you want them to be, rather than simply making do with the way things currently are" [BN1998]. Behaviors in this category include the ability to steer co-evolution and co-innovation—not merely taking a new path, but leading others towards playing the roles necessary to reach success [AK09].

We observe how what is generically referred to as cooperation and competition may appear in each of the above behavioral classes. As a consequence, they may differ considerably depending on the distinctive features of the class they originate in.

Behavior constitutes a first "axis" or category in our discussion of resilience in systems and organizations. As we have seen, similar behaviors may be exerted by individual and collective systems with different features and traits, ranging from bacteria to business organizations and including natural and computer-based systems. This leads to consider as a second important "axis" is our discussion what we refer to as an entity's individual and collective "Persona"—which is the subject of next subsection.

### INDIVIDUAL AND COLLECTIVE PERSONAE

Last section reviewed behaviors, viz. changes brought about by systems or organizations with respect to their environments. We now shift our attention to the actors of these changes and their individual and social identity.

A fundamental assumption in our discussion is borrowed from holonic organizations [Ko1967], actor-network theory [Lat1996], and common sense, and states that the nature of any individual system may always be regarded as a collective one. As well known, from the architectural and the functional points of view, with the exception of the most basic natural forms, any entity consists of subparts—that is any system is in fact an organization of constituent systems, or in other words a system-of-systems; however, such collective system can also be regarded for convenience as one individual (that is, a non-divisible entity) either because of its cumulative emerging active behavior (in the sense described in previous subsection), or because of its architecture, or for the strong identification of the individual with the society it belongs to. This concept is known as punctualization in actor-network theory. As an example, an organic system like the human body has a cohesive architecture with clearly identifiable boundaries (a body); its architecture is statically defined with distinct parts that take well-defined and permanent roles—for example, the autonomic nervous system—each of which is still a system-of-systems. The collection of these structural parts and their organizations constitute what we commonly refer to as the *physical person*, in this case a human being. Likewise, a company is a collection of sub-systems each assuming a certain role and providing some form of service and capital, and referred to as a whole as a *legal person*. Important differences here are the absence of a "physical body"—no clearly identifiable boundaries exist—and the fact that sub-systems are defined less statically (for instance, they may more easily change role or be replaced). A society of bees is yet another example of a system-of-systems with a loose architecture and no actual physical body, though the system may assume coherent and clearly identifiable physical forms and active behaviors such as those of a beehive or a swarm [Mae1909]. Bees may in fact be considered as a whole as an example of *social person*.

A final noteworthy example may be found in ancient Greek drama and provides us with possibly the first known representation of the concept of systems-of-systems. In fact, a classic role in Greek drama is that of the Chorus—a group of people representing the People's collective consciousness. To create the feeling of one collective system, members of the Chorus wear the same mask. The ancient Greek for mask is πρόσωπον (pronounced as "prosopon"), which later became *persona*, Latin for person, possibly also through the Etruscan *phersu*.

This leads to the second relationship in our treatise: In what follows we shall use the term "persona" to refer to the identity emerging from the social union of multiple cooperating sub-systems. Furthermore we shall use the term "personization" to refer to such punctualisation [Lat1996]. In the rest of this subsection we provide a classification of individual and collective personae.

Persona emergence may be transient, intermittent, or persistent, depending on the characteristics of the underlying social organization and the dynamic evolution of the inter-system relationships. An important property of personae is then identity robustness under specified or unforeseen perturbations, namely the second feature of resilience introduced in Section "Resilience". Identity robustness may be the result of a simple social organization (in this case the persona is said to be stable, that is, deterministic and modelizable as a simple dynamic system [Jen04, DF11]). Identity robustness may also characterize more complex organizations (structures able that is to switch between multiple strategic options such that changes are dynamically tolerated). We shall use the term "loss" to refer to a failure in the emergence of an intended persona. Losses may be characterized as transient (or temporary), intermittent, or permanent, depending on the situation at hand [La1995].

Most of the classes originally found in Boulding's [Bou1956] system-of-systems classifications, plus others inspired by more recent theories [CCHD05, Lee08, Zu10, DeB10], may be used to derive a new classification characterizing the common forms of personae and their resilience. In the current classification we distinguish:

- Servo-mechanisms, namely personae whose "movement" is fully pre-determined by external laws (e.g. the laws of physics). Here the organization is very simple and the persona is the most stable—sub-systems are assembled together so as to be (literally) the cogs of some tool or device designed specifically to serve a given purpose. Roles are fixed at design time. All actions are synchronous and strictly regulated by the movement of the other constituent cogs, as in meshing gears. The environment is assumed to be static and unchanging. Perception is therefore unnecessary, thus the perception system is absent or very limited [DF13c]. Identity persistence is only threatened by physical faults (e.g. internal or external changes due to worn-out parts, the application of external forces, or resource exhaustion). Terms such as "failure" are used to refer to persona losses.
- Simple control mechanisms like e.g. thermostats. As in servo-mechanisms, behavior here is a function of a fixed and predetermined design and of external laws, but in this case the mechanism is able to adjust its behavior teleologically (re: Subsection "Behaviors") tracking the value of some environmental figure (e.g., temperature). Typically such mechanisms operate along a single dimension and with no predictive behavior. Also in this case the persona is highly stable—the unitarian identity emerging from the multiple underlying activities may only be threatened by component failures. Also in this case we refer to persona losses as to failures.
- Simple "biological mechanisms" such as biological cells. The identity and the health status of a cell are mostly determined by its collection of proteins [Anon12a]—which brings to the foreground its collective nature. It is through the collaborative self-organization of proteins that cells fulfill their functions and exhibit teleological (though non-predictive) behaviors [Gus12]. Identity robustness is in this case the result of self-organization of simple constituents. The emerging persona is highly dependent on the robustness of the underlying components and the organization of their interactions. In this class, persona losses are usually referred to as "cellular disorders", "cellular diseases", or "cell deaths."
- Cyber-physical systems, i.e. open [Hey1998] embedded devices equipped with (i) sensors able to estimate physical quantities and (ii) actuators able to operate certain physical changes [Lee08]. Fast localization of peers and acquisition of knowledge is possible thanks to the emergence of new technology such as the Internet-of-Things [Fl10]. Software plays a major role in the emergence of the personae: Due to their characteristics, and in particular to software's ability to quickly evolve, cyber-physical systems are characterized by multiple personae that emerge in function of their interaction with human beings or their environments. Miniaturization makes it possible to easily deploy or even "spray" such devices [Za04], which fade into common human environments (ambient computing). Context awareness and situation identification [YeDM12] make it possible to track human-oriented activities and offer advanced personalized services. Autonomic behavior and in particular self-organization are common traits, though self-awareness and self-consciousness as well as the complex cognitive processes typical of man are (still) not present.
- "Plants," i.e. stationary systems comprising different complex specialized sub-systems. Also in this case the emerging persona is robust short of external events and negative environmental conditions (e.g. water deficit, parasites, fire, ...) The stationary system assumption strengthens the link with the deployment environment but simplifies the organization, e.g. reducing considerably the perception requirements [Ni10, DF13c]. Consequently, teleological behavior is very simple and lacks any form of proactiveness. Plant disease and plant death are common terms used to refer to persona losses in plants.
- "Animals," viz. complex mobile systems with extensive multimodal perception capabilities and simple forms of extrapolative behavior. A classic example of such behaviors is that of a cat pursuing a running mouse and moving towards its prey's extrapolated future position [RWB1943]. Competitive behaviors are commonly observed. A complex organization is required, which mandates the solution of several design trade-offs [Ni08, WTW1955]. The limited extrapolative ability is complemented by forms of innate behavior (instinct). This "design" results in a complex persona characterized by self-awareness and simple forms of self-consciousness. Despite these noteworthy features, animals commonly manifest limited individuality and their persona willingly collapses in "greater" social personae such as the flock, the pack, the anthill, or the beehive. Disease and death are common terms used with reference to persona losses in this class.
- "Human beings," viz. complex mobile systems characterized by extensive self-awareness, a high degree of self-consciousness, and high order extrapolative behaviors. Design trade-offs of natural evolution in this case privileged the ability to process information and derive knowledge from experience, as evidenced by the larger physical space allocated to cognitive functions. The persona emerging from these design choices is very complex and often characterized by a strong sense of the individual. As regrettably well known and already pointed out, the latter may produce conflicting and improvident behaviors that may scale up so as to jeopardize entire ecosystems [Ha1968]. Again, human diseases or death are terms used to refer to persona losses.
- Social organizations of complex mobile beings: *Networks*-of-systems in which "the unit [..] is not perhaps the person but the role—that part of the person which is [only] concerned with the organization or situation in question. Social organizations might be defined as a set of roles tied together with channels of communication" [Bou1956]. As already mentioned, this class is characterized by collections of loosely coupled, possibly sparse mobile components with no tight containment unit (that is, no physical body). We distinguish four sub-classes of social organizations:
    o Societies of "animals", namely the already mentioned collective systems based on commensalism (as e.g. in the flock, the pack, the anthill, and the beehive) and symbiosis (as e.g. in lichens) [AF1983]. Commensalism commonly involves animals of the same species and includes both cooperation and competition. Symbiosis is an interspecies relationship based on diverse requirements with respect to a common environment. It involves interspecies cooperation—also known as mutualism. In societies of animals the persona is mostly strong and robust—short of harsh environmental conditions—and is often able to reach a harmony or dynamic equilibrium with the deployment environment [WL04, K1902].
    o Societies of "human beings". A large variety of organizations exist in this case, including, e.g., the family, the clan, the guild, the community, the city, the state, and the business

organization. Complex forms of social interaction characterize this class. Self-interest is common, but so are also non-contractual social interactions and other-regarding behaviors [Bow06]. As pointed out in the cited reference, behaviors are also influenced by so-called "social preferences", which in some cases lead individuals to cooperate in spite of self-interested motives. Other noteworthy features of this class are the fast inter-communication and mobility of members, both of which are sustained by science and technology through accrued knowledge and experience. In the face of adverse environmental conditions the persona of human societies appears to be less capable than that of animal societies to maintain the already mentioned dynamic equilibrium with the environment [WL04, Ha1968]. This may be ascribed to the very cognitive and behavioral characteristics of its constituents as well as to cultural reasons (among others, the diffusion of individualism and utilitarianism [Sm1776]). The resulting overpopulation, pollution, and resource exhaustion, often coupled with "the logic of the commons" [Ha1968], may jeopardize in the long run not just the human society's persona but also that of the whole terrestrial ecosystem.
   - Systems of cyber-physical systems, i.e., networks of interacting cyber-physical systems. These societies share many of the traits of the above introduced class of cyber-physical systems including the ability to self-adapt and self-organize, high reconfigurability, context- and situation-awareness, as well as collaborative and competitive behaviors (opportunistic communication based on multi-user diversity [Vi06], resource competition [KSY12], altruism [WFK11], etc.) Emerging personae depend on the context and situation at hand. Collective forms of organizations, e.g. swarms [BCP10], are common.
   - Cyber-physical societies, namely "mixed societies" [CCHD05] of networked beings and systems encompassing "not only the cyber and physical spaces but also humans, knowledge, society and culture" [Zu10]. The emerging personae in this case are a function of a so-called "cyber-physical-physiological-psychological-socio-mental environment". Coupling the peculiar abilities and know-how's of the previous classes, such systems are expected to "operate autonomously in the real world through e.g. scene and context understanding, anticipation and reaction / adaptation to changes, manipulation and navigation, as well as symbiotic human-machine relations" [Anon12b]. A concept similar to cyber-physical societies and based on actor-network theory [Lat1996] is that of fractal social organizations [DF13b]. More information on the characteristics of such systems is provided in next section.
- Ecosystems, namely collectivities of "interacting organizations and individuals" [Moo1996] characterized by complex dynamic behaviors and advanced social models such as collective strategies, i.e. "the joint mobilization of resources and formulations of actions within collectivities of organizations" [AF1983]. The awareness of a "social energy" [DeB10] greater than the sum of each constituent's and the use of such energy to shape the landscape of future opportunities constitute the main reasons that bring organizations to define an ecosystem. Anexample is given by business ecosystems, which "can create a much larger and more valuable market than they ever could by working individually" [BN1998]. This class of personae is characterized by "purposive anticipatory behaviors", i.e. the last and most advanced class of behaviors presented in subsection "Behaviors". Such behaviors include, e.g., the already mentioned co-opetition, in which "a network of key players cooperate and compete with each other to create maximum profitability" in the face of a rapidly changing and highly turbulent environment. The collective nature of such strategy allows taking into account the "multiple and overlapping linkages among organizations" [AF1983] such as the current highly turbulent and technology-empowered environments. The resilience of ecosystems is subjected to their ability to create value for the participants, in turn dependent on technological and behavioral uncertainty [AK10].

A final aspect we would like to focus our attention on is the fact that persona creation and losses are also related to the relation between the individual and the greater self. Centripetal and centrifugal forces exist which lead individuals respectively to merge into the social persona and to escape from it. Such forces and especially their effect on human beings have been not only studied but also celebrated in works of art[5]. For the sake of brevity we will not discuss the nature and form of relationships that may be established by individuals within a collective.

We now shift our attention to the third category in our discussion of resilience: Organizational and control structures of social organization.

## ORGANIZATIONAL AND CONTROL STRUCTURES OF SOCIAL ORGANIZATION

In previous sections we introduced the two concepts of behavior (viz., any change of an entity with respect to its environment) and persona (namely individual identity as well as the identity emerging from the social union of multiple interrelated sub-systems or sub-organizations). Here we focus *on the way a social union is structured and organized.* "Partnership", "reference architecture", "organizational structure", "control structure", and simply "organization" are terms commonly used to refer to such concept [Ryu03]. In what follows we shall use the term "social organization", so as to emphasize the collective nature of the systems we are focusing on. The term social organization is in fact used in [Bou1956] to refer to the general class of systems-of-systems, which Boulding concisely defines as "a set of roles tied together with channels of communication". In the rest of this subsection our focus is on the way those roles are tied together.

A far from exhaustive classification of structures of social organization includes the following cases:

- Centralized control organizations. In this case a single focal entity takes charge of control of the whole system and maintains information of all ongoing activities. A distributed perception apparatus conveys environmental data to the focal entity, which is the sole responsible for defining a response strategy to maximize survivability and value capture. From the perspective of system personae (re: previous section), we remark how the focal entity is the core of the system and its personization: The persona of the system is in fact that of the focal entity, as e.g. the queen bee is the punctualization of the whole beehive. A positive trait of this form of organization is simplicity, though several shortcomings can also be observed:
    - The organization does not scale well [Ni08, Ni10]. In particular reactive behavior does not scale with number of connections, in that the focal entity needs to input and process sequentially all feedback information.
    - The organization relies on the health and conduct of a single entity (the focal entity is a so-called "single-point-of-failure" [La1995] and a single-point-of-congestion).
    - The organization is rather difficult to adjust and reconfigure [DF13b]; in particular adding a new entity calls for the introduction of a new communication line between that entity and the focal one.
    - Finally, as already pointed out by many scholars, it is doubtful that in the face of turbulent environmental conditions the focal entity could be able to singlehandedly monitor, analyze, and process all the ongoing events, including the interdependencies that rapidly manifest and evolve in such an environment [AF1983].

    In conclusion, this organization matches well to static and known environmental conditions and small-scale systems that do not often require reconfiguration.

- Hierarchical control organizations are characterized by a top-down flow of commands and a bottom-up flow of feedback information. As observed, e.g., in [Ryu03], these control/feedback flows are limited to the immediate subordinate/supervisor level. In such *teleological hierarchies* behaviors are translated into simpler and simpler actions while feedbacks are aggregated into more and more complex and abstract pieces of knowledge. The higher in the hierarchy, the more complex becomes the planning and the longest the planning horizon. The organization may then use in each layer of the hierarchy the entities that are best matching that layer's role.
A hierarchy scales better than a centralized organization, but it is not free of other shortcomings:
- Communication through nearest neighbors translates in a propagation delay. Information needs to visit all the intermediate layers to get to the top.
- The planning at higher levels may be disrupted by events at lower levels. For instance, if one intermediate layer sifts out or corrupts a certain piece of information, all the upper layers will not be aware of it.
- Intermediate supervisors typically have a narrower "task environment" [AF1983] which, while improving their efficiency, may well condition the quality of their action. Wrong deductions at one level ripple upward in the hierarchy and the accumulated error may lead the focal entity to the wrong decision path.
- Multiple distinct subordinate/supervisor relationships exist. This introduces extra complexity that needs to be known and taken into account when the system must be maintained, revised, optimized, repaired, or adapted.

---
[5] For instance, the "anguished dichotomy'" [Gur05] between the concept of individual and that of a "willing cog" in a greater mechanism is the "central conflict" [Wil05, p. 114] at the core of the play "Masse-Mensch" by Ernst Toller [Tol01]. Another example is the already cited "Cento Mani e Cento Occhi" [DGV1972].

The net result is that this class does not match well to turbulent environments.

- Heterarchical Control Organizations are those in which autonomous components, called tasks or agents, coexist in a flat structure without subordinate/supervisor relationships and with "considerable latitude and freedom to exercise strategic choice" [AF1983]. Resources are available to tasks and are allocated through a dynamic market mechanism based on competition. This makes the organization much more robust and tolerant of failures with respect to both the centralized and the hierarchical organizations. Simple rules and equal relationships govern the functions of this organization, which makes the system less complex and considerably easier to extend, modify, and adapt. Information is distributed so there is no single point of teleological planning and no global knowledge of the overall system or of the environmental conditions. As a consequence, central scheduling and resource planning are impractical. The emerging persona is not as predictable or easily identifiable as in previous organizations. Local behaviors are non-deterministic and global performance cannot be easily predicted.

- In an attempt to couple the benefits of hierarchical with those of heterarchical organizations several new "distributed organizations" have been proposed, all based on the idea that hierarchy and autonomy are not irreconcilable opposites. Other similarities in these organizations are the common origin (all of them are biologically-inspired); the assumption of the collective (i.e., composite) nature of systems; and the adoption of a single building block repeated at different scales of a hierarchy. Examples of organizations in this class are so-called bionic, holonic, and fractal organizations [DF13b]. An important aspect common to all such organizations lies in the fact that they all assume an organization to be structured as a network of nodes having the same function and repeated at different scales. Similarly to directories in a computer file system, organizations are built from the recurrence of a same "organizational building block". Such building block is called respectively modelon, holon, and fractal. Through this building block social organizations are composed of autonomous entities that "are simultaneously a part and a whole, a container and a contained, a controller and a controlled" [SSH00]. Obviously this trait matches well with the binary nature of individual and collective personae (re: previous section)—with the "self" being at the same time a sub-system and a super-system. Modelons, holons, and fractals operate through a set of configuration rules, called *canon*, which represents a "genetic code" of sort. Those structures are autonomous entities that establish cooperative relationships and are characterized by the emergence of stability, flexibility, and by efficient use of the available resources [DS13]. Among the peculiar differences between the three already mentioned bio-inspired organizations is the way the canon is defined. In particular in the modelon and the holon the canon is fixed while in the fractal dynamic restructuring and regrouping is not limited to the organization but also pertains to its canon, which is allowed to evolve. Examples of the above organizations are the Fractal Company [Wa1993], the Fractal Factory [TWN1998], Holonic Manufacturing Systems [SMHS1996], and Fractal Social Organizations [DF13b, DS13].

# DISCUSSION

In last section we introduced three categories for the discussion of resilience in systems and organizations. In the current section we consider three examples—three organizations for ambient assistance to the elderly and the impaired. For each organization we briefly highlight which behaviors, personae, and social organizations they exhibit. We show how, in some cases, a mismatch exists among the categories. Through a discussion we argue that said mismatch may explain the difference in quality and resilience in the three examples. A number of conjectures are finally put forward to associate our indicators with factors affecting the quality of resilience.

## CLASSIC SOLUTION

Traditional organizations for the assistance of the elderly and the impaired are experiencing a progressive transition towards unmanageability because of the steady increase of the share of the world's elderly and impaired population coupled with a structural shortage of caregivers [SDe10, FP12]. A classic organization in this category is health monitoring. Typically, health monitoring organizations deploy sensing devices to track the state of patients (for instance elderly or impaired persons) [FP12, LS13]. Those devices have a simple behavior: They constantly monitor a number of context figures describing the patients' conditions; figures are then logged and sent periodically to medical experts, e.g. on a daily basis. In addition, the devices check whether the registered figures can collectively describe a safe situation—as defined by a set of threshold figures. If that is not the case an alarm is triggered and a doctor is alerted. This constitutes a sort of "safety net" around the elderly or impaired. Safety net is the term we shall use in the current subsection to refer to this organization. We observe what follows:

- The personae involved in the above very simplified model of a safety net include the ones listed in Table 1.

| Social persona | Individual persona #1 | Individual persona #2 | Individual persona #3 | Individual persona #4 | Individual persona #5 |
|---|---|---|---|---|---|
| Safety net | Doctor | Cyber-physical things | Organs | Patient | … |
| Cyber-physical thing | Sensing devices | Localization devices | Communication devices | Store devices | … |
| Doctor | Medical doctor | Communication devices | Medical tools | Mobility assets | … |
| Communication devices | Phones | Pagers | Communication lines | … | … |
| Patient | Organ #1 | Condition #1 | Organ #2 | Condition #2 | … |

Table 1: Personae typically involved in a Safety Net.

- Social organization is hierarchical: A safety net is a doctor that reacts to feedbacks from cyber-physical things that react to feedbacks from the patient, as represented by a subset of his/her organs.
- Expected behaviors include, ordered from most complex to simplest, the extrapolative behaviors of doctors, the teleological behaviors of the cyber-physical "things", the active behavior or organs, and the passive behaviors of patients.

We highlight the adjective "expected" at the beginning of previous sentence. In fact the listed behaviors are not necessarily those *intrinsic* to the personae in question. In particular, patients here stand at the bottom of the behavioral hierarchy and in practice are equated to the role of objects. We call this a Behavior-vs-

Organization-vs-Persona mismatch (BOP). BOP mismatches such as this one are likely to impact severely on the involved people's dignity and life attitude [SDe10].

A few remarks are in order:

- The safety net organization does not exhibit very complex behaviors; no social strategy is ever formulated, e.g. only predefined forms of social collaboration are exploited. We conjecture that this may be a consequence of the rigid hierarchical social organization of the safety net.
- The typical shortcomings of hierarchies are experienced: Lack of scalability; dependence on the supervisor layers to react and take the right decision (single-point-of-failure); dependence on the subordinate layers to perceive and forward changes; lack of flexibility and adaptability; reaction delays; information propagation delays, etc [DS13].
- Due to our societies' structural shortage of professional caregivers, a same doctor is likely to serve multiple safety nets and multiple patients; this exacerbates reaction delays and the risks of correlated failures. In the face of a turbulent environment (due to e.g. the onset of a pandemic or a strong increase in the rate of the elderly in population) the time devoted to each patient may become insufficient to guarantee the system design goals.
- The cost of the professional caregiver is likely to impact non-negligibly on service costs.

Finally, it is important to point out how, by looking separately at the behavioral, systemic, and organizational aspects of this classic organization, we have been able to identify two related structural mismatches:

- First, we unraveled a discrepancy (a BOP) between an expected behavior and an intrinsic one. This reveals a possible role mismatch resulting in under-utilization of the available resources.
- Secondly, the vast majority of the players in the safety net organization—namely, the patients—do not take active part into the organization; in other words, they do not contribute with any active behavior to the emergence of the social persona. This is particularly severe given the scarcity of resources characterizing the environment.

We now discuss a second organization, the mutual assistance community, and show how this organization makes a better use of the available resources and better matches the characteristics of its intended environment.

## MUTUAL ASSISTANCE COMMUNITY

A Mutual Assistance Community (MAC) is a socio-technical system for the assistance of the elderly and the impaired [SDe07, SDe10, SDe13]. The MAC is based on three main concepts: Member, coordination center, and participant.

- The MAC member is either a human being or a cyber-physical thing that executes a protocol called "subscription". Subscribing is logically similar to logging into a computer system, but results in fact in requesting the admission into the MAC cyber-physical society. By subscribing the member publishes a number of pieces of information: Its *identity* (e.g. a name or a part number), *persona* (in the sense discussed in Section "Individual and Collective Personae"), *location* (continuously updated via a location tracking service), plus a *set of relationships* (e.g., member A is parent of member B) and a *set of service advertisements* (for instance member A's ability to play the role of "general practitioner", "nurse", "informal caregiver", or "accelerometer")—as defined in a custom ontology. Updates are published for dynamically changing figures. Members may also publish notifications: For instance a cyber-physical thing playing the role "accelerometer" may issue notification "member *x* has fallen", meaning that the person the accelerometer is watching is likely to have fallen. Notifications are also expressed in terms of a custom ontology and may be considerably more complex than in this example [SDe13].

- The MAC coordination center (CC) works as a "semantic switchboard" receiving and servicing all the available service advertisements and all the notifications published by its members. Each new submitted entry triggers a semantic match with all those related entries that are already known to the CC. If a satisfactory match can be found, the corresponding members are informed and associated servicing protocols are launched—for instance, a protocol to deal with one of the members falling in her/his house. If no match can be found the CC re-enters its main processing loop and awaits a new publication.

- Members enrolled for the execution of protocols are called "participants". Participants work in collaborative heterarchy, each one providing contributions as prescribed by the role it has assumed: For instance an informal caregiver located nearby may go and assess a situation *de visu*, thus providing a valuable "snapshot" of the situation to a participant professional caregiver still on his/her way; the latter may use the acquired information, e.g., to customize treatment or request specialized care. As a result of the dynamic enrollment, participants may come and go and correspondingly the heterarchy may change its composition and may shrink or grow in number. A formal way to represent this process is that of a random walk through the space of all possible participants in the current node. Figure 2 shows such space for a MAC consisting of six members.

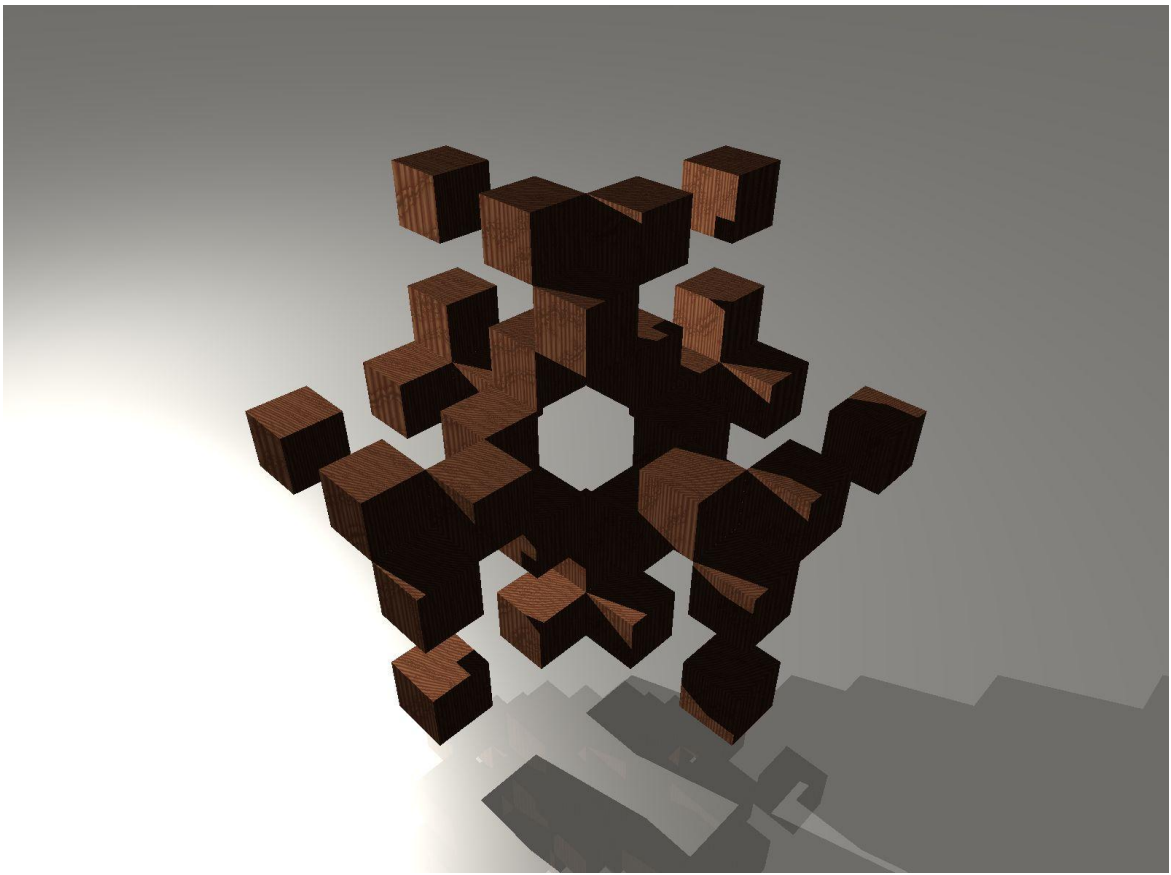

Figure 2. The picture shows the space of all possible "teams" of participants in a MAC with six members playing three roles. Role 1 and role 3 are played in this case by 1 member while role 2 by 4 members. More information on this may be found in [DF13b].

By considering the three categories introduced in the current paper we observe the following:

- The personae involved in a MAC that are more relevant in this discussion are those listed in Table 2.

| Social persona | Individual persona #1 | Individual persona #2 | Individual persona #3 | Individual persona #4 | Individual persona #5 |
|---|---|---|---|---|---|
| Mutual Assistance Community | Coordination center | Members | Cyber-physical things | Communication devices | |
| Member | Patients | Professional caregivers | Informal caregivers | Cyber-physical things | |
| Coordination center | Services infrastructure | Semantic matcher | Data storage | … | |
| Cyber-physical thing | Sensing devices | Localization devices | Communication devices | … | |
| Communication devices | Phones | Pagers | Communication lines | … | |

Table 2. Personae typically involved in a Mutual Assistance Community.

- Social organization is in this case a mixture of the centralized and the heterarchical: A MAC "is" a coordination center that reacts to feedbacks from all members by selectively relying information to a subset of members; once this selection is operated, a heterarchy is created to deal with the events.

- Expected behaviors are those intrinsic to the involved personae, as voluntarily selected by the involved personae. As a consequence, no structural limitation or BOP mismatch exists. What is probably more remarkable, the MAC does not assume any specific role of members—and in particular it does not assume that patients need always be on the receiving side. In fact we observed that, when notifications are not pertaining to alarming situation requiring professional assistance, patients may establish a mutual collaborative relationship in which each side provides some form of assistance to the other side [SDe07].

The above observations lead us to formulate the following remarks:

- By enabling the introduction of an abundant source of active behaviors, the "self-serve" model of the MAC not only decreases social costs but also allows a more turbulent environment to be addressed by making a more sensible use of the available resources.
- In so doing the MAC also promotes a healthier life style and fosters social contacts, physical and cognitive activity, and participation in society.
- By coupling the centralized and the heterarchical approach, the MAC overcomes some of the limitations of those social organizations (e.g. the impossibility of central scheduling and resource planning of heterarchies); still, some limitations cannot easily be overcome. In particular, the central coordinator is a single-point-of-failure and a single-point-of-congestion, and the behavior of participants may diverge from what expected. Though solutions exist to mitigate these deficiencies, it is still difficult to achieve quantitative guarantees of safety and timeliness.

We now provide and discuss the design of a third organization, the fractal social organization, and show how this new organization provides a structural improvement for some of the limitations of the MAC.

## FRACTAL SOCIAL ORGANIZATIONS

Fractal Social Organizations (FSO) may be concisely described as a generalization of the MAC described in previous section. Several major differences need to be pointed out. First and foremost, an FSO member is not restricted to being an individual system (e.g. a human being or a cyber-physical thing) but may also be, e.g., a MAC or even another FSO. This recursive definition translates into a hierarchy of layers. Each of said layers is organized similarly to a MAC, with a coordination center that is the personization of the whole layer. Within each layer, events such as "a member has fallen" may be detected and this leads to enrollments in a way similar to that of the MAC and presented in the previous section: roles necessary for the execution of servicing protocols are sought among those that may be played by the members in the MAC. A major difference in the enrollment of the FSO with respect to the MAC's is given by so-called exceptions: When a role cannot be found in the current layer, the FSO coordination center does not simply re-enter its processing loop but rather it propagates the event to the next level upward in the hierarchy. This goes on until the first suitable candidate participant is found or until some threshold is met. This creates an ad hoc, temporary MAC whose CC is elected and whose members are not restricted to a single layer but can span across multiple layers of the FSO. This MAC is called a Social Overlay Network (SON) and constitutes a temporary means for members situated at different layers to cooperate. The objective and lifespan of the newly spawn SON are determined by the triggered servicing protocol.

As it was the case for MAC, also in FSO enrollment is carried out via semantic matching. More information on this process may be found in [SDe13]. A formal description of the FSO activities, roles, and enrollment processes is out of the scope of this paper and may be found in [DF13b].

As we have done with the Safety Net and the MAC, here we formulate a number of remarks.

- A FSO for ambient assistance of the elderly and the impaired may include, e.g., the personae in Table 3.

| Social persona | Individual persona #1 | Individual persona #2 | Individual persona #3 | Individual persona #4 | Individual persona #5 |
|---|---|---|---|---|---|
| Fractal Social Organization | Fractal Social Organizations | Mutual Assistance Communities | Hospitals / healthcare organization | Smart-house | … |
| Mutual Assistance Community | Members | Coordination center | Cyber-physical things | Communication devices | … |
| Hospital | Care units | Transport units | Communication units | Administration units | … |

Table 3. Personae typically involved in a FSO for the ambient assistance of the elderly or the impaired.

As mentioned above, FSO members may be "simultaneously a part and a whole, a container and a contained, a controller and a controlled" [SSH00], namely the FSO has a fractal organization. This property is expressed in the first row of the above table: A FSO is first and foremost a collection of other FSO's, which in the domain of ambient assistance may take the shape of mutual assistance communities, hospitals and other healthcare organizations, smart-houses, professional and informal caregivers, patients, etc.

- The FSO is not limited t*o* a single domain but may span through several ones. It is then in principle possible that a FSO to deal with e.g. civil protection and one to deal with e.g. social assistance cooperate together to respond to some drastic and unexpected environmental shock.
- The global social organization of a FSO is fractal, but within a FSO any other social organization is possible; for instance, hospitals are typically hierarchically organized, while MAC's are hybrid organizations exhibiting centralized and heterarchical traits.

As in the MAC, the FSO does not impose any structural limitation of personae, which contribute according to their peculiar nature, business, and mission. Thus no BOP mismatch is introduced. Furthermore, the presence of multiple, self-elected coordination centers eliminates the single-point-of-failure deficiency of the MAC, while the availability of multiple FSO instances allows multiple redundant "planes" of action to be enacted in parallel, which may be beneficial to enhance predictability and resilience.

A partial implementation of the FSO is currently being carried out in the framework of "Little Sister", a Flemish project financed by iMinds and the Flemish Government Agency for Innovation by Science and

Technology (IWT). Little Sister aims at delivering a low-cost telemonitoring solution for home care [LS13]. Preliminary evaluation shows that the FSO outperforms the MAC in efficiency, scalability, and maintainability [DS13,DF13c]. Other studies highlighted in the FSO the spontaneous emergence of self-similarity and modularity [DF13b]—two properties that have been conjectured by other researchers as prerequisites to evolvability and resilience [Clu13,Wag96].

## OBSERVATIONS AND LESSONS LEARNED

A major observation we can draw when considering the above examples is that, at least in some cases, collective systems exhibit what we have called as "BOP mismatches". This occurs, for instance, when an organization prohibits or discourages the behaviors typical of a participating persona. An example of this is, e.g., a patient served by a safety net and unable to exercise the complex behaviors typical of the "human being" persona. Another case of BOP mismatch is discussed in [CKL08] in the context of the social response exhibited by New Orleans to Hurricane Katrina: the authors observed how a variety of "shadow responders" emerged spontaneously and remarked how

> ""Emergent" individuals or organizations that respond to unaddressed needs are characteristic of all disaster responses. In responding to Katrina, they were **sometimes refused or poorly used** by government officials [..] These "shadow responders" often emerge from households, friends and family, neighborhoods, non-governmental and voluntary organizations, businesses, and industry. In New Orleans, we estimate that they provided most of the initial evacuation capacity, sheltering, feeding, health care, and rebuilding, and much of the search and rescue, cleanup, and post-Katrina funding. These individuals and organizations **would have been able to do more if the tri-level system (city, state, federal) of emergency response was able to effectively use, collaborate with, and coordinate the combined public and private efforts**. How to do so, in advance of hazard events, is a central task of enhancing community resilience."

After consider the above examples and facts, in what follows we put forward a number of conjectures.

1. We conjecture that BOP mismatches may contribute to "centrifugal forces" leading to the dissolution of the social persona: The collective system thus "breaks down" into a set of uncorrelated individual constituents. As identity persistence is a pre-requisite to resilience (cf. Section "Resilience"), the resulting loss of identity translates in a loss of quality of resilience.
2. The more complex the constituents' personae, the stronger will be the centrifugal forces associated with BOP mismatches. Rationale for this conjecture is that complex personae are characterized by self-awareness, self-consciousness as well as with stronger individualistic features. Because of this, the lack of explicit "returns" for the individual would lead it to breaking down its relationship with the greater system.
3. On the contrary, the more evident and explicit are the individual advantages resulting from joining the "greater self", the stronger will be the "centripetal force" attracting individuals and cementing their social union; and the less pronounced will be the effect of possible BOP mismatches.

   Extreme environmental conditions [Mey11] due to, e.g., natural or man-induced disasters, may make in some case those advantages evident and explicit. The excellent response exhibited by the New Orleans community after Hurricane Katrina [CKL08] may be interpreted as a consequence of said phenomenon. The social response (i.e., the behaviors of the social entity) after Katrina was more effective in that the individuals melded more willingly into the "greater self" because the individual returns were more evident to everybody (e.g., the expectancy of survival or recovery for themselves and those in their "closer circles"[7]). As remarked above, an even better response would have been possible should the crisis management organization be able to avoid BOP mismatches by promoting and orchestrating the behaviors of the individual constituents.

---

[7] New Orleans's Mayor, Mr. Landrieu remarks how "The thing you learn [from Hurricane Katrina] is *there are a lot of folks are going through the same thing, and you can help each other*." http://www.businessweek.com/articles/2013-08-06/how-new-orleans-became-a-model-of-urban-resilience.

Conversely, when the advantages of a collective response are canceled, for instance, by the turbulence or the instability of the environments, the social persona collapses into that of its individual constituents, as well captured by the vernacular "Every man for himself."

## CONCLUSIONS

We discussed quality of resilience in collective systems. As a way to overcome the hard coupling between resilience and the environment our systems are set to operate in, we proposed to evaluate quality of resilience through a set of indicators: The enacted behaviors; the characteristics of the involved individual and social systems; and the employed organizational structures. For each of these attributes we provided a detailed classification and characterization. By considering a particular domain and three different organizations we have shown the presence of mismatches between the above three attributes. Similar mismatches have been observed in the social response to a well-known natural catastrophe. As a result of our observations we put forward a number of conjectures about the centripetal and centrifugal forces that affect the quality of social resilience. In particular we highlighted 1) the role of organization as an enabler or disabler of individual behaviors and personae; 2) the role of the persona and its behaviors as a trigger to social identity persistence; and 3) the role of the context and turbulence as a catalysts of either centripetal or centrifugal behaviors.

Our analyses also confirm that overcoming simplistic organization and purely individualistic and competition-oriented behaviors leads (at least in the treated cases) to a better exploitation of the available resources and to "social energy" [DB10]. Paraphrasing John Donne, our results show that "no system or organization is an island"[8] in the troubled sea of modern resource-scarce and technologically empowered environments. Realizing the "social persona" of resilience as a multi-attribute discipline including both individual as social aspects and influenced by centripetal and centrifugal forces may help, we conjecture, engineer design techniques and systematic approaches to evolve organizations so as to match variable environmental conditions. This in turn could help shifting the thresholds of unmanageability; avoiding or at least postponing the onset of chaotic behaviors, and capturing value.

Our current work is focused towards the study of collective resilience strategies and the development, simulation, and testing of our fractal social organizations [DF13b,DS13]. Our major and final conjecture put forward here is that the fractal organization of our FSO would make it possible to integrate in a natural way the responses to crisis enacted by "shadow responders" (informal actors) with those of pre-existing organizations such as the U.S. "tri-level system of emergency response" [CLK08], thus creating an effective response to the need "**to effectively use, collaborate with, and coordinate the combined public and private efforts**". That need was referred to in the cited reference as "a central task of enhancing community resilience".

*ACKNOWLEDGMENTS*
This work was partially supported by iMinds—Interdisciplinary institute for Technology, a research institute funded by the Flemish Government—as well as by the Flemish Government Agency for Innovation by Science and Technology (IWT).

## REFERENCES

---

[8] "No man is an island, Entire of itself. Each is a piece of the continent, A part of the main". From John Donne, "No Man Is An Island".

[AK10] Ron Adner and Rahul Kapoor, "Value creation in innovation ecosystems: How the structure of technological interdependence affects firm performance in new technology generations", Strategic Management Journal, Vol. 31, pp. 306–333, 2010

[Anon12a] Anonymous, "Protein Function", in Scitable—A Collaborative Learning Space for Science, Nature Publishing Group: 2012. URL: http://www.nature.com/scitable/topicpage/protein-function-14123348

[Anon12b] Anonymous, "FP7-Information and Communications Technologies Work Programme 2013", European Commission, 2012

[AF1983] W. Graham Astley and Charles J. Fombrun, "Collective Strategy: Social Ecology of Organizational Environments", The Academy of Management Review, Vol. 8, No. 4 (Oct., 1983), pp. 576–587

[Avi1995] A. Avizienis, "The Methodology of N-Version Programming". In Lyu, M., ed.: Software Fault Tolerance. John Wiley & Sons (1995) 23–46.

[BCP10] Eric Bonabeau, David Corne, and Riccardo Poli, "Swarm intelligence: the state of the art", Natural Computing, Vol. 9, No.3, 2010, pp. 655–657, 2010, DOI: 10.1007/s11047-009-9172-6

[Bou1956] Kenneth E. Boulding, "General Systems Theory—The Skeleton of Science", Management Science, Vol. 2, No. 3 (April 1956), pp. 197–208

[Bow06] Samuel Bowles, "Microeconomics: Behavior, Institutions, and Evolution", Princeton University Press, 2006, ISBN: 0-691-09163-3

[BN1998] Adam Brandenburger, Barry Nalebuff, "Co-opetition—A Revolutionary Mindset that Combines Competition and Cooperation", Doubleday, 1998, ISBN: 9780385479509

[CCHD05] Gilles Caprari, Re Colot, Jos Halloy, and Jean-Louis Deneubourg, "Building Mixed Societies of Animals and Robots", IEEE Robotics & Automation Magazine, Vol. 12, pp. 58–65, 2005

[Car+12] L. Carlson et al. Resilience: Theory and Applications. Tech. Report Argonne Nat. Lab. ANL/DIS-12-1, 2012.

[Clu13] Clune, Jeff, Mouret, Jean-Baptiste, & Lipson, Hod.. The evolutionary origins of modularity. In Proceedings of the Royal Society b: Biological sciences, 280(1755), 2013.

[CKL08] Colten, C. E., Kates, R.W., and Laska, S.B. Community Resilience: Lessons from New Orleans and Hurricane Katrina. CARRI Research Report 3. Available online at http://www.resilientus.org/wp-content/uploads/2013/03/FINAL_COLTEN_9-25-08_1223482263.pdf) (Sept. 2008)

[DFB10] Vincenzo De Florio and Chris Blondia, "Service-oriented communities: Visions and contributions towards social organizations," in Proc. of the On the Move to Meaningful Internet Systems: OTM 2010 Workshops, ser. Lecture Notes in Computer Science, R. Meersman, T. Dillon, and P. Herrero, Eds., Springer Berlin / Heidelberg, 2010, Vol. 6428, pp. 319–328, DOI: 10.1007/978-3-642-16961-8_51

[DF13a] Vincenzo De Florio, On the Constituent Attributes of Software and Organisational Resilience, Interdisciplinary Science Reviews, Vol. 38, No. 2, Maney Publishing, 2013.

[DF13b] Vincenzo De Florio, Mohamed Bakhouya, Antonio Coronato, and Giovanna Di Marzo, "Models and Concepts for Socio-technical Complex Systems: Towards Fractal Social Organizations", Systems Research and Behavioral Science, Vol. 30, No. 6, Wiley & Sons, Nov. 2013.

[DF13c] De Florio, V. Preliminary Contributions Towards Auto-Resilience. In Proceedings of the 5th International Workshop on Software Engineering for Resilient Systems (SERENE 2013), Lecture Notes in Computer Science Vol. 8166. Springer (2013) 141–155.

[DS13] De Florio, Hong Sun, Jonas Buys, and Chris Blondia, "On the Impact of Fractal Organization on the Performance of Socio-technical Systems", in Proc. of the 2013 International Workshop on Intelligent Techniques for Ubiquitous Systems (ITUS 2013), Vietri, Italy, 2013. IEEE.

[DGV1972] Francesco di Giacomo, Vittorio Nocenzi, "Cento Mani e Cento Occhi", in "Darwin", Ricordi, 1972

[Dr1981] Melvin Dresher, "The mathematics of games of strategy", Dover Publications Inc., New York, 1981. Theory and applications; Reprint of the 1961 original

[DSMF75] Edgar Dunn, Donald A. Schon, Ruth P. Mack and Andreas Faludi, "A Review Forum", Journal of the American Institute of Planners, Vol. 43, No.3, pp. 214-219, 1975, DOI: 10.1080/01944367508977880

[EaKl10] D. Easley and J. Kleinberg. "Networks, Crowds, and Markets: Reasoning about a Highly Connected World". Cambridge University Press, 2010

[ECW05] Saul Eisen, Jeanne Cherbeneau, and Christopher G. Worley, "A future-responsive perspective for competent OD practice". In W. G. Rothwell & R. Sullivan (Eds.), Practicing organization development (pp. 188-208). New York: Wiley-Pfeiffer, 2005

[Eug09] Eugster, P. T., Garbinato, B., and Holzer, A. Middleware support for context aware applications. In Garbinato, B., Miranda, H., and Rodrigues, L., eds. Middleware for Network Eccentric and Mobile Applications. Springer (2009) 305–322.

[Farr02] D. Farr. Indicator Species. In: Encycl. of Environmetrics. Wiley (2002)

[FHHNS07] Lee Feigenbaum, Ivan Herman, Tonya Hongsermeier, Eric Neumann and Susie Stephens, "The Semantic Web in Action", Scientific American, December 2007, URL: http://www.thefigtrees.net/lee/sw/sciam/semantic-web-in-action

[Fl10] Elgar Fleisch, "What is the Internet of Things? An Economic Perspective", Auto-ID Labs White Paper No. WP-BIZAPP-053, January 2010, URL: http://www.autoidlabs.org/uploads/media/AUTOIDLABS-WP-BIZAPP-53.pdf

[GPT06] Gaël Gueguen, Estelle Pellegrin-Boucher, and Olivier Torres. "Between cooperation and competition: the benefits of collective strategies within business ecosystems. The example of the software industry". EIASM 2nd Workshop on Co-opetition Strategy, Milan, Italy, 14–15 Sept., 2006, URL: http://www.gaelgueguen.fr/wp-content/uploads/GueguenPellegrinTorres.pdf

[Gur05] A. E. Gurganus, "Sarah Sonja Lerch, née Rabinowitz: The Sonja Irene L. of Toller's *Masse-Mensch*." In German Studies Review, Vol. 28, No. 3, pp. 607–220. October 2005

[Gus12] Ksenia Guseva, "Formation and Cooperative Behavior of Protein Complexes on the Cell Membrane", Springer:2012, Springer Theses, ISBN 978-3-642-23987-8. Originally published as a doctoral thesis of the Institute of Complex Systems and Mathematical Biology of the University of Aberdeen, UK

[Ha1968] Garrett Hardin, "The Tragedy of the Commons", Science, Vol. 162 (Dec. 1968), pp. 1243–1248, DOI: 10.1126/science.162.3859.1243

[Hey1998] F. Heylighen, "Basic concepts of the systems approach," in Principia Cybernetica Web, F. Heylighen, C. Joslyn, and V. Turchin, Eds. Principia Cybernetica, Brussels, 1998. [Online]. Available: http://pespmc1.vub.ac.be/SYSAPPR.html

[HP06] John L. Hennessy and David A Patterson, "Computer Architecture: A Quantitative Approach", 4th edition, Morgan Kaufmann, San Francisco, 2006

[Jen04] Erica Jen, "Stable or robust? What's the difference?", in E. Jen, editor, "Robust Design: a repertoire of biological, ecological, and engineering case studies", SFI Studies in the Sciences of Complexity, pages 7–20. Oxford University Press, 2004

[John1989] Johnson, B. W. Design and Analysis of Fault-Tolerant Digital Systems. Addison-Wesley (1989)

[KSY12] Valerie King, Jared Saia, and Maxwell Young, "Resource-Competitive Communication", Computing Research Repository, Vol. abs/1202.6456, URL: http://arxiv.org/abs/1202.6456

[Ko1967] Arthur Koestler, "The Ghost in the Machine". 1967. London: Arcana Books, 1989

[La1995] J.-C. Laprie, "Dependability—Its Attributes, Impairments and Means". In B. Randell, J.-C. Laprie, H. Kopetz, & B. Littlewood (Eds.), Predictably Dependable Computing Systems (pp. 3–18). Berlin: Springer, 1995

[Lap05] J.-C. Laprie et al. "Resilience for the scalability of dependability". Proc. ISNCA 2005, pp. 5-6.

[Lat1995] B. Latour. On actor-network theory. a few clarifications plus more than a few complications. Soziale welt, 47 : 369–381. 1996.

[Laws1986] Aristotle, H. Lawson-Tancred, H. De Anima (On the Soul). Penguin (1986)

[Lee08] Edward A. Lee, "Cyber Physical Systems: Design Challenges", University of California, Berkeley, EECS Department, Technical Report No. UCB/EECS-2008-8, January 23, 2008, URL: http://www.eecs.berkeley.edu/Pubs/TechRpts/2008/EECS-2008-8.pdf

[LS13] Little Sister Consortium. Little Sister: low-cost monitoring for care and retail. Project description. Available online at http://www.iminds.be/en/research/overview-projects/p/detail/littlesister (2013)

[Mae1909] Maurice Maeterlinck, "The Life of the Bee", New York: Dodd, Mead & Co., 1909. Translated by Alfred Sutro from "La vie des abeilles", 1901

[Mar06] Marshall, Charles R. Explaining the Cambrian "Explosion" of Animals. Annual Review of Earth and Planetary Sciences, Vol. 34. Annual Reviwes (2006): 355-384. DOI: 10.1146/annurev.earth.33.031504.103001.

[Mey09] J.F. Meyer, Defining and evaluating resilience: A performability perspective. In: Proc. Int.l Work. on Performability Modeling of Comp. & Comm. Sys. (2009)

[Mey11] R. A. Meyers (Ed.) "Extreme Environmental Events – Complexity in Forecasting and Early Warning". Springer (2011)

[Mi1973] Donald N. Michael, "Learning to Plan and Planning to Learn: The Social Psychology of Changing Toward Future-Responsive Societal Learning", San Francisco: Jossey-Bass Publishers, 1973

[Moo1996] James E. Moore, "The Death of Competition: Leadership and Strategy in the Age of Business Ecosystems", Harper Business, 1996

[Ni08] Thomy Nilsson, "Solving the Sensory Information Bottleneck to Central Processing in Complex Systems". In A. Yang and Y. Shan (eds.), "Intelligent Complex Adaptive Systems", pp. 159–186. IGI Global:2008, Hershey, PA

[Ni10] Thomy Nilsson, "How neural branching solved an information bottleneck opening the way to smart life", in Proceedings of the 10th International Conference on Cognitive and Neural Systems, Boston University, MA

[RaXu95] Randell, Brian and Xu, Jie. The Evolution of the Recovery Block Concept. In Lyu, Michael. Software Fault Tolerance. John Wiley & Sons (1995) 1 – 21.

[RWB1943] Arturo Rosenblueth, Norbert Wiener and Julian Bigelow, "Behavior, Purpose and Teleology". Philosophy of Science, 10(1943), S. 18–24

[Ryu03] Kwangyeol Ryu, "Fractal-based Reference Model for Self-reconfigurable Manufacturing Systems", Doctoral dissertation, Pohang University of Science and Technology, Korea, 2003

[Sa1995] Joe Sachs, "Aristotle's Physics: A Guided Study". Rutgers University Press, Masterworks of Discovery, 1995, ISBN: 0-8135-2192-0

[SaMa11] Salles, Ronaldo M. and Marino, Donato A. Strategies and Metric for Resilience in Computer Networks. The Computer Journal (2011). DOI: 10.1093/comjnl/bxr110.

[SWBO09] D. Schultz, P. G. Wolynes, E. Ben Jacob and J. N. Onuchic, "Deciding Fate in Adverse Times: Sporulation and Competence in *Bacillus subtilis*", Proc. Nat.l Acad. Sci. Vol. 106, pp. 21027–21034, 2009

[Sim08] L. Simoncini, "Resilience assessment and dependability benchmarking: challenges of prediction," DSN Workshop on Resilience Assessment and Dependability Benchmarking, 2008.
[Sm1776] Adam Smith, "An Inquiry into the Nature and Causes of the Wealth of Nations", London : J.M. Dent & Sons ; New York : E.P. Dutton, 1910, URL: http://archive.org/details/wealthofnationss01smituoft
[SSH00] P. Sousa, N. Silva, T. Heikkila, M. Kallingbaum and P. Valcknears, P. "Aspects of Co-operation in Distributed Manufacturing Systems". Studies in Informatics and Control Journal, 9 (2). 2000, pp. 89–110
[SMHS1996] Sugimura, N., Moriwaki, T., Hozumi, K. and Shinohara, Y., Modeling of holonic manufacturing system and its application to real-time scheduling. Mfg. Sys., 1996,25, 345–352
[SDe07] Hong Sun, Vincenzo De Florio, Ning Gui, and Chris Blondia, "Participant: A New Concept for Optimally Assisting the Elder People", In Proc. of the 20th IEEE Int.l Symp. on Computer-Based Medical Systems (CBMS-2007), IEEE Comp. Soc: June 2007
[SDe10] Hong Sun, Vincenzo De Florio, Ning Gui, and Chris Blondia, "The Missing Ones: Key Ingredients Towards Effective Ambient Assisted Living Systems", Journal of Ambient Intelligence and Smart Environments, Vol. 2, No. 2, April 2010, pp. 109–120
[SDe13] Hong Sun, Vincenzo De Florio, and Chris Blondia, "Implementing a Role Based Mutual Assistance Community with Semantic Service Description and Matching", in Proc. of the Int.l Conference on Management of Emergent Digital EcoSystems (MEDES), ACM: October 2013.
[Tal12] Taleb, Nassim Nicholas. Antifragile: Things That Gain from Disorder. Random House Publishing Group: 2012.
[TWN1998] A. Tharumarajah, A. Wells, and L. Nemes, "Comparison of emerging manufacturing concepts," in Proc. of the IEEE International Conference on Systems, Man, and Cybernetics, Vol. 1, Oct 1998, pp. 325–331
[Tol01] Ernst Toller, "Masse-Mensch". In "Toller Plays: vol. 1 (Transformation, Masses Man, Hoppla We're Alive!)", 2001: Oberon Books Ltd.
[TriDG09] K.S. Trivedi, K. Dong Seong, and R. Ghosh. Resilience in computer systems and networks. Computer-Aided Design - Digest of Technical Papers, 2009. ICCAD 2009. IEEE/ACM International Conference on, pp.74,77, 2-5 Nov. 2009
[vdS1999] W.H. van der Schalie, et al. Animals as sentinels of human health hazards of environmental chemicals. Environ. Health Persp. 107(4) (1999)
[WFK11] M. Waibel, D. Floreano, and L. Keller, "A Quantitative Test of Hamilton's Rule for the Evolution of Altruism", PLoS Biol 9(5): e1000615, 2011, DOI: 10.1371/journal.pbio.1000615
[Wag96] Wagner, G. and Altenberg, L. 1996a. Perspective: complex adaptations and the evolution of evolvability. Evolution, 50:967–976.
[Wa1993] H. Warnecke and M. Hüser, "The Fractal Company: A Revolution in Corporate Culture", Springer, 1993
[WL04] C. T. Webb and S. A. Levin, "Cross-system perspectives on the ecology and evolution of resilience", in Jen, E. (ed.), "Robust Design: a repertoire of biological, ecological, and engineering case studies", SFI Studies in the Sciences of Complexity, pp. 151–172, Oxford University Press:2004
[WTW1955] G. Weddell, D. Taylor, and C. Williams, "The Patterned Arrangement of Spinal Nerves to the Rabbit Ear". J. Anat., 89, pp. 317–342, 1955.
[Wil05] P. Willett, "The Cutting Edge of German Expressionism: The woodcut novel of Frans Masereel and its influences." In N. H. Donahue (Ed.), "A Companion to the Literature of German Expressionism", Studies in German Literature, Linguistics, and Culture. 2005: Camden House.
[YeDM12] J. Ye and S. Dobson and S. McKeever. "Situation Identification Techniques in Pervasive Computing: A Review". Pervasive and Mobile Computing, Vol. 8, No.1, pp. 36–66, Elsevier: 2012, DOI: 10.1016/j.pmcj.2011.01.004
[Za04] F. Zambonelli, M.-P. Gleizes, M. Mamei, and R. Tolksdorf. "Spray computers: frontiers of self-organization for pervasive computing". In Proc. of the 13th IEEE International Workshops on Enabling Technologies: Infrastructure for Collaborative Enterprises, 2004, pp. 403–408, DOI: 10.1109/ENABL.2004.58
[Zu10] H. Zhuge, "Cyber Physical Society—A Cross-Disciplinary Science". On-line document. URL: http://www.knowledgegrid.net/~h.zhuge/CPS.htm